\magnification \magstep1
\raggedbottom
\openup 2\jot 
\voffset6truemm
\def\cstok#1{\leavevmode\thinspace\hbox{\vrule\vtop{\vbox{\hrule\kern1pt
\hbox{\vphantom{\tt/}\thinspace{\tt#1}\thinspace}}
\kern1pt\hrule}\vrule}\thinspace}
\centerline {\bf NON-LOCALITY AND ELLIPTICITY IN A}
\centerline {\bf GAUGE-INVARIANT QUANTIZATION}
\vskip 1cm
\leftline {Giampiero Esposito and Cosimo Stornaiolo}
\vskip 0.3cm
\noindent
{\it Istituto Nazionale di Fisica Nucleare, Sezione di Napoli,
Mostra d'Oltremare Padiglione 20, 80125 Napoli, Italy}
\vskip 0.3cm
\noindent
{\it Universit\`a degli Studi di Napoli Federico II, Dipartimento
di Scienze Fisiche, Complesso Universitario di Monte S. Angelo,
Via Cintia Edificio G, 80126 Napoli, Italy}
\vskip 1cm
\noindent
{\bf Abstract}.
The quantum theory of a free particle in two dimensions with 
non-local boundary conditions on a circle is known to lead to
surface and bulk states. Such a scheme is here generalized to
the quantized Maxwell field, subject to mixed boundary conditions.
If the Robin sector is modified by the addition of a
pseudo-differential boundary operator, gauge-invariant boundary
conditions are obtained at the price of dealing with 
gauge-field and ghost operators which become pseudo-differential.
A good elliptic theory is then obtained if the kernel occurring in
the boundary operator obeys certain summability conditions, and
it leads to a peculiar form of the asymptotic expansion of the
symbol. The cases of ghost operator of negative and positive order
are studied within this framework.
\vskip 100cm
\leftline {\bf 1. Introduction}
\vskip 0.3cm
\noindent
In the late eighties, non-local boundary conditions for operators 
of Laplace type found some interesting applications to physical
problems, i.e. the behaviour of a free Bose gas and the phenomenon
of Bose--Einstein condensation. More precisely, the work in Ref. 1
studied the stationary Schr\"{o}dinger equation for a free particle
in polar coordinates on a circle of radius $R$:
$$
-\left[{\partial^{2}\over \partial r^{2}}
+{1\over r}{\partial \over \partial r}
+{1\over r^{2}}{\partial^{2}\over \partial \varphi^{2}}
\right]u(r,\varphi)=Eu(r,\varphi) 
\; \; r < R,
\eqno (1.1)
$$
subject to the boundary condition
$$
\left . {\partial u \over \partial r} \right |_{r=R}
+R \int_{-\pi}^{\pi}q_{R}(R(\varphi-\theta))u(R,\theta)d\theta=0.
\eqno (1.2)
$$
In Eq. (1.2), $q_{R}$ is defined as
$$
q_{R}(x) \equiv {1\over 2\pi R}\sum_{l=-\infty}^{\infty}
e^{ilx/R} \int_{-\infty}^{\infty}
e^{-ily/R}q(y)dy,
\eqno (1.3)
$$
where $q$ is integrable and square-integrable on ${\bf R}$:
$q \in L_{1}({\bf R}) \cap L_{2}({\bf R})$. Both positive ($E>0$) and 
negative ($E<0$) eigenvalues are admissible. In the former case,
two sets of eigenfunctions are obtained: surface states, which
decrease exponentially with increasing distance from the boundary,
and bulk states, corresponding to more extended eigenfunctions.
For negative eigenvalues, only surface states are found to occur.

Interestingly, (even) for a free Schr\"{o}dinger operator, a 
non-local boundary condition such as the one in (1.2) may lead to
two sets of solutions which, in other physical problems, are
normally obtained under quite different conditions. We have
therefore tried to understand whether the above scheme admits 
an extension to some fundamental field theory. The first
non-trivial example is given, in our opinion, by an Abelian
gauge theory, i.e. the free Maxwell field 
$F_{ab} \equiv \nabla_{a}A_{b}-\nabla_{b}A_{a}$ in vacuum,
where $A_{b}$ is the electromagnetic potential and $\nabla$ 
is the Levi--Civita connection of the background geometry. At the
classical level, this may be studied by imposing a supplementary
condition of Lorenz type:
$$
\nabla^{b}A_{b}=0,
\eqno (1.4)
$$
so as to obtain the homogeneous wave equation for the electromagnetic
potential. In the quantum theory via path integrals, one performs
Gaussian a\-ve\-ra\-ges o\-ver ga\-uge 
func\-tio\-nals,$^{2}$ he\-re\-af\-ter de\-no\-ted by
$\Phi(A)$, to avoid ``summing'' over gauge-equivalent field
configurations. This is achieved by adding the term 
${[\Phi(A)]^{2}\over 2\alpha}$ to the original Maxwell Lagrangian,
where $\alpha$ is a gauge parameter. In particular, in the one-loop
semiclassical theory, the resulting gauge-field operator 
$P_{a}^{\; b}$ acting on $A_{b}$ is found to have a non-degenerate
leading symbol. This operator is hence elliptic, with a well
defined Green function, thanks to the introduction of a term that
``breaks'' the gauge invariance properties of ${1\over 4}F_{ab}F^{ab}$.
Moreover, to the gauge functions of the classical theory, for which
$$
{ }^{f}A_{b} \equiv A_{b}+\nabla_{b}f
\eqno (1.5)
$$
leads to the same field equations as $A_{b}$, there correspond two
fermionic ghost fields in the quantum theory (usually referred to
as ghost and anti-ghost, although they are actually independent$^{2}$),
with the associated ghost operator. Its ``classical counterpart''
is clearly obtained if one remarks that the supplementary condition
(1.4) is preserved under the action of (1.5) if and only if the
gauge function obeys the equation $\cstok{\ }f=0$. The form of the
equation obeyed by $f$ will depend, of course, on which
supplementary condition is chosen.

If one accepts the view that the potential $A$ is more fundamental
than the Maxwell field $F=dA$ (this is suggested by the 
Aharonov--Bohm effect and by the emphasis on differential 
operators in the path integral), the boundary conditions should 
also involve the potential and the ghost in the first place
(rather than using components of $\vec E$ and $\vec B$). This
formulation, although not compelling, is certainly appropriate
if one studies the gauge-field operator $P_{a}^{\; b}$, since
this acts on $A_{b}$, and therefore cannot be properly studied
without specifying the boundary conditions on $A_{b}$. A set of
gauge-invariant boundary conditions is obtained upon requiring 
that the tangential components of $A$ should 
vanish at the boundary:
$$
[A_{k}]_{\partial M}=0,
\eqno (1.6)
$$
jointly with the gauge-averaging functional, i.e.
$$
[\Phi(A)]_{\partial M}=0.
\eqno (1.7)
$$
One can in fact prove that both Eq. (1.6) and Eq. (1.7) are 
preserved under gauge transformations on the potential if the
ghost field obeys homogeneous Dirichlet conditions  
(see Ref. 3 and our Appendix):
$$
[\varepsilon]_{\partial M}=0.
\eqno (1.8)
$$
In particular, if (free) Maxwell theory is quantized in the
Lorenz gauge (cf. Eq. (1.4)), equations (1.6) and (1.7) are
found to imply a Robin boundary condition on the normal component
of the potential, i.e.
$$
\left[{\partial A_{0}\over \partial n}
+A_{0}{\rm Tr}K \right]_{\partial M}=0,
\eqno (1.9)
$$
where $K$ is the extrinsic-curvature tensor of the boundary.

Section 2 studies a non-local modification of Eq. (1.9) inspired
by Eq. (1.2), jointly with the request of gauge invariance of the
whole set of boundary conditions. The resulting gauge-field and
ghost operators are studied in Sec. 3, and the conditions for
an elliptic theory are analyzed in Sec. 4. Concluding remarks
are presented in Sec. 5, and relevant details are given
in the Appendix.
\vskip 0.3cm
\leftline {\bf 2. Boundary Conditions: Non-Locality and Gauge Invariance}
\vskip 0.3cm
\noindent
Since we are interested in a generalization of 
the model described by Eqs. (1.1)--(1.3) to
Maxwell theory expressed in terms of the potential, we are led to
modify the Robin sector of the boundary conditions (1.6) and 
(1.9), by requiring that
$$
\left[{\partial A_{0}\over \partial n}
+A_{0}{\rm Tr}K \right]_{\partial M}
+R \int_{-\pi}^{\pi}q_{R}(R(\varphi-\theta))A_{0}(R,\theta)
d\theta=0.
\eqno (2.1)
$$
However, to avoid having a non-local boundary operator which spoils
gauge invariance of the boundary conditions, we should be able to
regard Eq. (2.1) as a particular case of Eq. (1.7) when the boundary
condition (1.6) is imposed. Thus, upon choosing the normal to the
boundary in the form $N^{b}=(1,0)$, we are led to consider a
gauge-averaging functional (hereafter all indices take the values
$0$ and $1$)
$$
\Phi(A) \equiv \nabla^{b}A_{b}+N^{b}Q_{b}, 
\eqno (2.2)
$$
where $Q_{a}$ is defined by
$$
Q_{a} \equiv N_{a}r \int_{-\pi}^{\pi}q_{r}(r(\varphi-\theta))
N^{b}A_{b}(r,\theta)d\theta,
\eqno (2.3)
$$
which ensures that (see (2.1))
$$
[N^{b}Q_{b}]_{\partial M}=[Q_{0}]_{\partial M}
=R \int_{-\pi}^{\pi}q_{R}(R(\varphi-\theta))
A_{0}(R,\theta)d\theta.
\eqno (2.4)
$$
The full set of boundary 
conditions is now given by (1.6), (2.1) and the Dirichlet
condition (1.8) on the ghost.
\vskip 0.3cm
\leftline {\bf 3. Ghost and Gauge-Field Operators}
\vskip 0.3cm
\noindent
The ghost operator is obtained by taking the difference between
the gauge-averaging functional $\Phi(A)$ and the same functional
when acting on the gauge-transformed potential 
${ }^{\varepsilon}A_{b} \equiv A_{b}+\nabla_{b}\varepsilon$
(see (1.5)). In our problem, by virtue of the choice (2.2),
one finds
$$
\Phi(A)-\Phi({ }^{\varepsilon}A)={\cal P}\varepsilon,
\eqno (3.1)
$$
where the action of $\cal P$, the ghost operator, reads
$$
{\cal P} \varepsilon =-\cstok{\ }\varepsilon
-r \int_{-\pi}^{\pi}q_{r}(r(\varphi-\theta))N^{c}\nabla_{c}
\varepsilon(r,\theta)d\theta ,
\eqno (3.2)
$$
having defined $\cstok{\ } \equiv g^{ac}\nabla_{a}\nabla_{c}
=\nabla^{b}\nabla_{b}$. Such a ghost operator is an
integro-differential operator by virtue of the occurrence 
of $Q_{b}$ in (2.2).

The corresponding gauge-field operator is also integro-differential, 
because it is obtained from the ``Lagrangian''
$$
L \equiv {1\over 4}F_{ab}F^{ab}+{[\Phi(A)]^{2}\over 2\alpha},
\eqno (3.3)
$$
after writing it in the form
$$
L={1\over 2}A^{a} \; P_{a}^{\; b} \; A_{b} + 
{\rm total} \; {\rm derivative}.
\eqno (3.4)
$$
For this purpose we use the identity
$$
{1\over 4}F_{ab}F^{ab}={1\over 2}(\nabla_{a}A_{b})
(\nabla^{a}A^{b})-{1\over 2}(\nabla_{a}A_{b})
(\nabla^{b}A^{a}),
\eqno (3.5)
$$
the Leibniz rule and the commutator of covariant derivatives to
prove that the first term on the right-hand side of (3.3) contributes
$$
S_{a}^{\; b} \equiv -\delta_{a}^{\; b}\cstok{\ }
+\nabla_{a}\nabla^{b}+R_{a}^{\; b}
\eqno (3.6)
$$
to $P_{a}^{\; b}$. As is well known, this operator has a degenerate
leading symbol$^{4}$ and hence is not invertible. To deal with the second
term on the right-hand side of (3.3) we use (2.2) and the identities
$$ 
{1\over 2\alpha}[\Phi(A)]^{2}={1\over 2\alpha}
(\nabla^{b}A_{b})(\nabla^{c}A_{c})+{1\over \alpha}
(\nabla^{b}A_{b})(N^{c}Q_{c}) 
+{1\over 2\alpha}(N^{b}Q_{b})(N^{c}Q_{c}),
\eqno (3.7)
$$
$$
(\nabla^{b}A_{b})(\nabla^{c}A_{c})=\nabla^{b}(A_{b}\nabla^{c}A_{c})
-A^{a}\nabla_{a}\nabla^{b}A_{b},
\eqno (3.8)
$$
$$ 
(\nabla^{b}A_{b})(N^{c}Q_{c})=\nabla^{b}(A_{b}N^{c}Q_{c})
-A^{a}\nabla_{a} r \int_{-\pi}^{\pi}
q_{r}(r(\varphi-\theta))N^{b}A_{b}(r,\theta)d\theta.
\eqno (3.9)
$$
By virtue of (3.3)--(3.9), the gauge-field operator takes 
the form
$$ 
P_{a}^{\; b}=-\delta_{a}^{\; b}\cstok{\ }
+ \left(1-{1\over \alpha}\right)\nabla_{a}\nabla^{b}
+R_{a}^{\; b}+{1\over \alpha} T_{a}^{\; b} 
+{1\over \alpha}U_{a}^{\; b},
\eqno (3.10)
$$
where $T_{a}^{\; b}$ and $U_{a}^{\; b}$ are integral operators
defined by
$$
\Bigr(T_{a}^{\; b}A_{b}\Bigr)(r,\varphi) \equiv
-2 \nabla_{a} r \int_{-\pi}^{\pi}q_{r}(r(\varphi-\theta))
N^{b}A_{b}(r,\theta)d\theta ,
\eqno (3.11)
$$
$$ \eqalignno{
\; & \Bigr(A^{a}U_{a}^{\; b}A_{b}\Bigr)(r,\varphi) \equiv
\int_{-\pi}^{\pi}\int_{-\pi}^{\pi}A^{a}(r,\theta)r^{2}
q_{r}(r(\varphi-\theta))q_{r}(r(\varphi-\theta')) \cr 
& \; \; \; \; \; \; \; \; \; \; \; \; \; \; \; \; \; \; \; \;
\; \; \; \; \; \; \; \; \; \; 
\; \; \; \; \; \; \; \; \; \; \; \; \; \;
N_{a}N^{b}A_{b}(r,\theta')d\theta d\theta'.
&(3.12)\cr}
$$
\vskip 0.3cm
\leftline {\bf 4. Ellipticity}
\vskip 0.3cm
\noindent
The compatibility of gauge invariance of the boundary conditions
with their non-local character has been shown to lead to
integro-differential gauge-field and ghost operators. Before we
can regard all this as a viable scheme, some consistency checks
are in order. In particular, we are here interested in preserving
the ellipticity of the theory, which is known to hold when local
boundary conditions of mixed nature are imposed.$^{4}$

For this purpose, we here focus on the ghost operator given in (3.2),
because the novel features arising from the pseudo-differential
framework are already clear at that stage.
This makes it necessary to use the definition of ellipticity for
pseudo-differential operators, which is first given on open subsets
of ${\bf R}^{m}$ and then extended to deal with changes of coordinates,
as is shown in Secs. 1.3.1 and 1.3.2 of Ref. 4. The key steps are
as follows.$^{4}$
\vskip 0.3cm
\noindent
(i) A linear partial differential operator $P$ of order $d$ can be
written in the form
$$
P \equiv \sum_{|\alpha| \leq d} a_{\alpha}(x)D_{x}^{\alpha},
\eqno (4.1)
$$
where ($i$ denotes, as usual, the imaginary unit)
$$
|\alpha| \equiv \sum_{k=1}^{m}\alpha_{k},
\eqno (4.2)
$$
$$
D_{x}^{\alpha} \equiv (-i)^{|\alpha|}
{\left({\partial \over \partial x_{1}}\right)}^{\alpha_{1}}
...
{\left({\partial \over \partial x_{m}}\right)}^{\alpha_{m}},
\eqno (4.3)
$$
and $a_{\alpha}$ is a $C^{\infty}$ function on ${\bf R}^{m}$ for
all $\alpha$. The associated {\it symbol} is, by definition,
$$
p(x,\xi) \equiv \sum_{|\alpha| \leq d}a_{\alpha}(x)\xi^{\alpha},
\eqno (4.4)
$$
i.e. it is obtained by replacing the differential operator 
$D_{x}^{\alpha}$ by the monomial $\xi^{\alpha}$. The pair 
$(x,\xi)$ may be viewed as defining a point of the cotangent 
bundle of ${\bf R}^{m}$, and the action of $P$ on the elements 
of the Schwarz space $\cal S$ of smooth complex-valued 
functions on ${\bf R}^{m}$ of rapid decrease is given by
$$
Pf(x) \equiv \int e^{i(x-y)\cdot \xi}p(x,\xi)f(y)dy d\xi,
\eqno (4.5)
$$
where the $dy=dy_{1}...dy_{m}$ and $d\xi=d\xi_{1}...d\xi_{m}$ orders
of integration cannot be interchanged, since the integral is not
absolutely convergent.
\vskip 0.3cm
\noindent
(ii) Pseudo-differential operators are instead a more general class
of operators whose symbol need not be a polynomial but has suitable 
regularity properties. More precisely, let $S^{d}$ be the set of
all symbols $p(x,\xi)$ such that$^{4}$
\vskip 0.3cm
\noindent
(1) $p$ is smooth in $(x,\xi)$, with compact $x$ support.
\vskip 0.3cm
\noindent
(2) For all $(\alpha,\beta)$, there exist constants 
$C_{\alpha,\beta}$ for which
$$
\left | D_{x}^{\alpha}D_{\xi}^{\beta}p(x,\xi) \right |
\leq C_{\alpha,\beta} (1+|\xi|)^{d-|\beta|},
\eqno (4.6)
$$
for some {\it real} (not necessarily positive) 
value of $d$, where $|\beta| \equiv
\sum_{k=1}^{m} \beta_{k}$ (see (4.2)). The associated 
{\it pseudo-differential operator}, defined on the Schwarz space
and taking values in the set of smooth functions on ${\bf R}^{m}$
with compact support:
$$
P: {\cal S} \rightarrow C_{c}^{\infty}({\bf R}^{m}),
$$
is defined in a way formally analogous to Eq. (4.5).
\vskip 0.3cm
\noindent
(iii) Let now $U$ be an open subset with compact closure in
${\bf R}^{m}$, and consider an open subset $U_{1}$ whose closure
${\overline U}_{1}$ is properly included into $U$: 
${\overline U}_{1} \subset U$. If $p$ is a symbol of order $d$
on $U$, it is said to be {\it elliptic} on $U_{1}$ if there exists
an open set $U_{2}$ which contains ${\overline U}_{1}$ and
positive constants $C_{i}$ so that
$$
|p(x,\xi)|^{-1} \leq C_{1} (1+|\xi|)^{-d},
\eqno (4.7)
$$
for $|\xi| \geq C_{0}$ and $x \in U_{2}$,$^{4}$ where
$$
|\xi| \equiv \sqrt{g^{ab}(x)\xi_{a}\xi_{b}}
=\sqrt{\sum_{k=1}^{m} \xi_{k}^{2}}.
\eqno (4.8)
$$
The corresponding operator $P$ is then elliptic.
\vskip 0.3cm
\noindent
In our problem, we revert to the use of Cartesian coordinates, so
that the above definitions can be immediately applied. Hereafter,
$(x,y)$ and $(x',y')$ are coordinates of the points $X$ and $X'$
of ${\bf R}^{2}$, respectively. $Q_{a}$ is the operator defined
by a convolution (cf. Eq. (2.3))
$$
Q_{a}f(x,y) \equiv N_{a} \int_{-\infty}^{\infty} 
\int_{-\infty}^{\infty} Q(x-x',y-y')f(x',y') dx' dy',
\eqno (4.9)
$$
where the unit normal to the circle has components
$$
N_{1}=N_{x}={x\over \sqrt{x^{2}+y^{2}}},
\eqno (4.10)
$$
$$
N_{2}=N_{y}={y\over \sqrt{x^{2}+y^{2}}}.
\eqno (4.11)
$$
The gauge-averaging functional is taken to be (cf. Eq. (2.2))
$$ \eqalignno{
\; & \Phi(A) \equiv \nabla^{b}A_{b}+N^{b}Q_{b} \cr
&=\nabla^{b}A_{b}+\int_{-\infty}^{\infty}
\int_{-\infty}^{\infty}Q(x-x',y-y')N^{c}A_{c}(x',y')dx'dy'.
&(4.12)\cr}
$$
Hence the action of the ghost operator reads (cf. Eq. (3.2))
$$
{\cal P}\varepsilon=-\cstok{\ }\varepsilon
-\int_{-\infty}^{\infty} \int_{-\infty}^{\infty} Q(x-x',y-y')
N^{c}\nabla_{c}\varepsilon(x',y')dx' dy'.
\eqno (4.13)
$$
Recall now that the boundary conditions (1.6) and (1.7) are
gauge-invariant if and only if the ghost field obeys homogeneous
Dirichlet conditions (1.8) on the boundary. We can therefore use
the identity ($K$ is again the 
extrinsic-curvature tensor of the boundary) 
$$
QN^{c}\nabla_{c}\varepsilon=\nabla_{c}(N^{c}Q \varepsilon)
-({\rm Tr}K)Q \varepsilon -N^{c}(\nabla_{c}Q)\varepsilon,
\eqno (4.14)
$$
the divergence theorem (here
$B_{R} \equiv \left \{ x,y: x^{2}+y^{2} \leq R^{2} \right \}$):
$$
\int_{B_{R}}\nabla_{c}(N^{c}Q\varepsilon)
=\int_{\partial B_{R}} N^{c}Q \varepsilon \; d\sigma_{c},
\eqno (4.15)
$$
and integration by parts, to cast Eq. (4.13) in the form
$$
{\cal P}\varepsilon =-\cstok{\ }\varepsilon 
+\int_{-\infty}^{\infty} \int_{-\infty}^{\infty}
\Bigr[(({\rm Tr}K)+N^{c}\nabla_{c})Q \Bigr]
\varepsilon(x',y')dx' dy'.
\eqno (4.16)
$$
Equation (4.16) shows clearly that, on setting
$$
W(x,y) \equiv {\rm Tr}K+N^{c}\nabla_{c},
\eqno (4.17)
$$
the integral operator on the right-hand side has kernel given by
$$ \eqalignno{
\; & \chi(x,y;x',y') \equiv W(x,y) Q(x-x',y-y') \cr
&=({\rm Tr}K)Q({\widetilde x},{\widetilde y})
+N_{x}{\partial Q \over \partial {\widetilde x}}
+N_{y}{\partial Q \over \partial {\widetilde y}},
&(4.18)\cr}
$$
where ${\widetilde x} \equiv x-x', {\widetilde y} \equiv y-y'$.
Thus, the corresponding symbol can be evaluated from Eq. 
(2.1.36) in Ref. 5:
$$
p(x,\xi)=\int e^{-i z \cdot \xi}\chi(x;x-z)dz.
\eqno (4.19)
$$
In our case, Eqs. (4.18) and (4.19) imply that the symbol of the
ghost operator $\cal P$ in Eq. (4.16) is given by
$$ \eqalignno{ 
\; & p(x,y;\xi_{1},\xi_{2})=|\xi|^{2}+\int_{-\infty}^{\infty}
\int_{-\infty}^{\infty}e^{-i(z_{1}\xi_{1}+z_{2}\xi_{2})}
\chi(x,y;x-z_{1},y-z_{2})dz_{1}dz_{2} \cr
&=|\xi|^{2}+\int_{-\infty}^{\infty} \int_{-\infty}^{\infty}
e^{-i(z_{1}\xi_{1}+z_{2}\xi_{2})} F(x,y,z_{1},z_{2})dz_{1}dz_{2},
&(4.20)\cr}
$$
where 
$$
F(x,y,z_{1},z_{2}) \equiv \left(({\rm Tr}K)Q(z_{1},z_{2})
+N_{x} \left . {\partial Q \over \partial {\widetilde x}}
\right |_{z_{1},z_{2}}
+N_{y} \left . {\partial Q \over \partial {\widetilde y}}
\right |_{z_{1},z_{2}} \right),
\eqno (4.21)
$$
because ${\widetilde x}=z_{1}$ when $x'=x-z_{1}$, and
${\widetilde y}=z_{2}$ when $y'=y-z_{1}$.
To achieve ellipticity in the interior of $B_{R}$ we now
impose the majorization (4.7), re-expressed in the form
$$
|p(x,y;\xi_{1},\xi_{2})| \geq {\widetilde C}_{1}
(1+|\xi|)^{d},
\eqno (4.22)
$$
for all $|\xi| \geq C_{0}$, where ${\widetilde C}_{1} \equiv
C_{1}^{-1}$. On the other hand, by virtue of (4.20),
it is always true that
$$
|p(x,y;\xi_{1},\xi_{2})| \geq \left( |\xi|^{2}
- \left | \int_{-\infty}^{\infty} \int_{-\infty}^{\infty}
e^{-i(z_{1}\xi_{1}+z_{2}\xi_{2})} F(x,y,z_{1},z_{2})dz_{1}dz_{2}
\right | \right).
\eqno (4.23)
$$
To ensure ellipticity in $B_{R}- \partial B_{R}$
it is therefore sufficient to impose that
$$
{\widetilde C}_{1}(1+|\xi|)^{d} \leq \left( |\xi|^{2}
-\int_{-\infty}^{\infty} \int_{-\infty}^{\infty}
|F(x,y,z_{1},z_{2})| dz_{1}dz_{2}\right),
\eqno (4.24)
$$
where we have used the majorization
$$ \eqalignno{
\; & \left | \int_{-\infty}^{\infty} \int_{-\infty}^{\infty}
e^{-i(z_{1}\xi_{1}+z_{2}\xi_{2})}F(x,y,z_{1},z_{2})dz_{1}dz_{2}
\right | \cr
& \leq \int_{-\infty}^{\infty} \int_{-\infty}^{\infty}
|F(x,y,z_{1},z_{2})| dz_{1}dz_{2},
&(4.25)\cr}
$$
to go from (4.23) to (4.24).

For example, if $d < 0$, the majorization 
(4.24) can lead, for $|\xi| \geq C_{0}$, to
$$
C_{0}^{2}-{\widetilde C}_{1} (1+C_{0})^{d} \geq 
{\rm sup}_{x,y \in B_{R}}
\int_{-\infty}^{\infty} \int_{-\infty}^{\infty}
|F(x,y,z_{1},z_{2})| dz_{1} dz_{2},
\eqno (4.26)
$$
which is satisfied if
$$
Q(z_{1},z_{2}) \in L^{1}({\bf R}^{2}),
\eqno (4.27)
$$
and also (see (4.21))
$$
\left . {\partial Q \over \partial {\widetilde x}}
\right |_{z_{1},z_{2}} \; {\rm and} \;
\left . {\partial Q \over \partial {\widetilde y}}
\right |_{z_{1},z_{2}} \; 
\in L^{1}({\bf R}^{2}).
\eqno (4.28)
$$

In particular, the equality of left- and right-hand side can be
considered in (4.26), so that the resulting 
order $d$ of the ghost operator
$\cal P$ can be evaluated in the form
$$
d={\log \Bigr(C_{1}(C_{0}^{2}-I)\Bigr) \over \log(1+C_{0})},
\eqno (4.29)
$$
where
$$
I \equiv {\rm sup}_{x,y \in B_{R}} 
\int_{-\infty}^{\infty} \int_{-\infty}^{\infty}
|F(x,y,z_{1},z_{2})| dz_{1}dz_{2}.
\eqno (4.30)
$$
Thus, this particular order of the ghost operator is negative, in
agreement with the assumption leading to (4.26),
if $C_{1}(C_{0}^{2}-I) \in ]0,1[$.

Moreover, strong ellipticity should also be studied.$^{4,5}$ 
For this purpose, following Ref. 5, we assume that the symbol 
of the ghost operator given in Eq. (4.20) is {\it polyhomogeneous},
in that it admits an asymptotic expansion of the form
$$
p(x,y;\xi_{1},\xi_{2}) \sim \sum_{l=0}^{\infty}
p_{d-l}(x,y;\xi_{1},\xi_{2}),
\eqno (4.31)
$$
where each term $p_{d-l}$ has the homogeneity property
$$
p_{d-l}(x,y;t\xi_{1},t\xi_{2})
=t^{d-l}p_{d-l}(x,y;\xi_{1},\xi_{2}),
\eqno (4.32)
$$
for $t \geq 1$ and $|\xi| \geq 1$. The {\it principal symbol}
$p^{0}$ of the ghost operator is then, by definition,
$$
p^{0}(x,y;\xi_{1},\xi_{2}) \equiv p_{d}(x,y;\xi_{1},\xi_{2}).
\eqno (4.33)
$$
{\it Strong ellipticity} (see comments in Sec. 5) 
is formulated in terms of the principal
symbol, because it requires that$^{5}$
$$
{\rm Re} \; p^{0}(x,y;\xi_{1},\xi_{2})={1\over 2}\Bigr[
p^{0}(x,y;\xi_{1},\xi_{2})+p^{0}(x,y;\xi_{1},\xi_{2})^{*}
\Bigr] \geq c(x)|\xi|^{d}I,
\eqno (4.34)
$$
where $x \in B_{R}, c(x) > 0$ and $|\xi| \geq 1$. In other words,
given a positive function $c$, the product $c(x) |\xi|^{d} I$ should
be always majorized by the real part of the principal symbol 
of the ghost operator. Indeed, the symbol (4.20) is such that
$$ \eqalignno{
\; & p(x,y;t\xi_{1},t\xi_{2})=t^{2} 
\Bigr(\xi_{1}^{2}+\xi_{2}^{2}\Bigr) \cr
&+t^{-2}\int_{-\infty}^{\infty}\int_{-\infty}^{\infty}
e^{-i(z_{1}\xi_{1}+z_{2}\xi_{2})}F \left(x,y,{z_{1}\over t},
{z_{2}\over t} \right) dz_{1} dz_{2}.
&(4.35)\cr}
$$
By virtue of (4.31), (4.32) and (4.35) we find that the kernel $Q$
should have an asymptotic expansion such that
$$ \eqalignno{
\; & t^{2} |\xi|^{2}+t^{-2}\int_{-\infty}^{\infty}
\int_{-\infty}^{\infty} e^{-i(z_{1}\xi_{1}+z_{2}\xi_{2})}
F \left(x,y,{z_{1}\over t},{z_{2}\over t}\right)dz_{1}dz_{2} \cr
& \sim \sum_{l=0}^{\infty} t^{d-l}p_{d-l}(x,y;\xi_{1},\xi_{2}).
&(4.36)\cr}
$$
Moreover, the term on the right-hand side of (4.36) with $l=0$
should be the one occurring in the condition (4.34) for strong
ellipticity. Our understanding of the necessary class of kernels
has therefore made progress.

Last, if the order of the ghost operator $\cal P$ is positive and
even, one can use Theorem 1.7.2 in Ref. 5 to prove ellipticity
with Dirichlet boundary conditions as in Eq. (1.8). Since we have
previously discussed the case of negative order, it is necessary
to describe how such a positive order can be obtained. To begin
note that, if $f$ is a $C^{\infty}$ function on ${\bf R}^{m}$
with compact support, one can define a symbol of order $d$ for
any $d \in {\bf R}$ by using the formula$^{4}$
$$
p(x,\xi)\equiv f(x)(1+|\xi|^{2})^{d\over 2}.
\eqno (4.37)
$$
The associated kernel is then$^{5}$
$$
\chi(x,y)=(2\pi)^{-m}\int e^{i(x-y)\cdot \xi}p(x,\xi)d\xi.
\eqno (4.38)
$$
Consider now, for simplicity, the case $m=1$ and $d=4$. Such 
formulae make it then necessary to evaluate the kernel $\chi$ by 
studying the integral (hereafter $X \equiv x-y$)
$$
J(X) \equiv \int_{-\infty}^{\infty}e^{iX \xi}
(1+2\xi^{2}+\xi^{4})d\xi,
\eqno (4.39)
$$
which is meaningful within the framework of Fourier transforms
of distributions.$^{6}$ To get an explicit representation, we
write $J(X)$ in the form
$$
J(X) \equiv \int_{-\infty}^{\infty} \lim_{a \to 0}
e^{-a \xi^{2}}e^{iX\xi} (1+2\xi^{2}+\xi^{4})d\xi,
\eqno (4.40)
$$
and hence consider the parameter-dependent integrals
$$
J_{1,a}(X) \equiv \int_{-\infty}^{\infty}e^{-a\xi^{2}}
e^{iX\xi}d\xi=\sqrt{\pi \over a}e^{-X^{2}/4a},
\eqno (4.41)
$$
$$
J_{2,a}(X) \equiv 2\int_{-\infty}^{\infty}\xi^{2}e^{-a\xi^{2}}
e^{iX\xi}d\xi={1\over a}\sqrt{\pi \over a}e^{-X^{2}/4a} 
\left(1-{X^{2}\over 2a}\right),
\eqno (4.42)
$$
$$ \eqalignno{
\; & J_{3,a}(X) \equiv \int_{-\infty}^{\infty}\xi^{4}
e^{-a \xi^{2}}e^{iX\xi}d\xi \cr
&={3\over 4a^{2}} \sqrt{\pi \over a}e^{-X^{2}/4a} 
\left[1-{X^{2}\over a}+{1\over 12}{X^{4}\over a^{2}}\right].
&(4.43)\cr}
$$
The link with the theory of distributions is now clear if one
bears in mind that the following regular distribution:
$$
{\nu \over \sqrt{\pi}}e^{-\nu^{2}\tau^{2}}
$$
converges in the space of all continuous linear functionals
to the delta functional as $\nu \rightarrow \infty$. If one sets
$\nu={1\over \sqrt{a}},X=2\tau$, the integral $J_{1,a}$ converges
therefore to $\pi$ times the delta functional as $a \rightarrow 0$.
One can treat similarly $J_{2,a}$ and $J_{3,a}$, and hence prove
explicitly the distributional nature of the kernel $\chi$ obtained
from (4.37)--(4.39).

To sum up, positive orders of the ghost operator, which are
necessary to prove ellipticity with Dirichlet boundary conditions,
lead to a kernel $\chi(x,y;x',y')$ (see (4.18)) of distributional
nature. An explicit representation can be obtained by an
$m$-dimensional generalization of the integrals (4.41)--(4.43).
Moreover, the kernel $Q(x-x',y-y')$ which contributes to
$\chi(x,y;x',y')$ should be such that the asymptotic expansion
(4.36) holds. This makes it possible to pick out the principal
symbol which occurs in the strong ellipticity condition (4.36).
\vskip 0.3cm
\leftline {\bf 5. Concluding Remarks}
\vskip 0.3cm
\noindent
In the first part of our paper we have shown how to choose non-local
boundary conditions for the quantized Maxwell field in a way 
compatible with the request of complete gauge invariance of the 
resulting boundary operator. This scheme has been found to lead to
gauge-field and ghost operators of integro-differential nature.
In the second part of our paper we have studied more carefully the
ghost operator within the framework of pseudo-differential 
operators on ${\bf R}^{2}$. Interestingly, such an operator remains
elliptic in the interior of the region considered therein 
provided that the kernel occurring in the boundary operator
fulfills the summability conditions (4.27) and (4.28). Moreover,
strong ellipticity for the ghost holds if (4.36) and (4.34)
are satisfied. The above results are, to our
knowledge, completely new in the physical literature, although some 
non-local aspects in the quantization of Maxwell theory had been
studied, for example, in Refs. 7 and 8. The ultimate meaning of
strong ellipticity is that it ensures the existence of the
asymptotic expansion of the $L^{2}$-trace of the heat semigroup 
associated to the given operator,$^{4}$ so that the resulting
conformal anomaly is well defined in one-loop quantum theory.
From the mathematical point of view, strong ellipticity is a
precise formulation of existence and uniqueness of smooth solutions
with given boundary conditions in an elliptic boundary-value
problem.$^{4}$

As far as physics is concerned, it now appears crucial to understand
which novel features of quantized gauge theories can result from
the consideration of pseudo-differential boundary-value problems.
Last, but not least, such investigations might have a non-trivial
impact on the attempts of quantizing the gravitational field, where
the role of a non-local formulation$^{9,10}$ is also receiving 
careful consideration.
\vskip 5cm
\leftline {\bf Appendix}
\vskip 0.3cm
\noindent
Since not all readers might be familiar with boundary conditions 
on the potential and the ghost field for Maxwell theory, we find
it appropriate to prove why the ghost has to obey homogeneous
Dirichlet conditions in our problem.

If tangential components of the potential are set to zero at the
boundary as in Eq. (1.6), the preservation of this part of the
boundary conditions under gauge transformations leads to
$$
[\partial_{k}\varepsilon]_{\partial M}=0.
\eqno ({\rm A}.1)
$$
But tangential derivatives only act on the part of $\varepsilon$ 
depending on the local coordinates on the $(m-1)$-dimensional
boundary, and hence, if Eq. (1.8) is imposed, Eq. (A.1) is
automatically satisfied. In other words, the operations of
tangential derivative and restriction to the boundary turn
out to commute. 

Moreover, if $\cal P$ is symmetric and elliptic, the field
$\varepsilon$ can be expanded in a complete orthonormal set
of $C^{\infty}$ eigenvectors $\varepsilon_{\lambda}$ of $\cal P$,
for which
$$
{\cal P}\varepsilon_{\lambda}=\lambda \varepsilon_{\lambda}.
\eqno ({\rm A}.2)
$$
In other words, one can write
$$
\varepsilon=\sum_{\lambda}C_{\lambda}\varepsilon_{\lambda},
\eqno ({\rm A}.3)
$$
which implies, by virtue of Eq. (3.1),
$$
\Phi(A)-\Phi({ }^{\varepsilon}A)
=\sum_{\lambda}\lambda C_{\lambda} \varepsilon_{\lambda}.
\eqno ({\rm A}.4)
$$
Thus, if Eq. (1.8) holds, which is satisfied when
$$
[\varepsilon_{\lambda}]_{\partial M}=0 \; \; 
\forall \; \lambda,
\eqno ({\rm A}.5)
$$
then the vanishing of the gauge-averaging functional at the
boundary is a gauge-invariant boundary condition as well on the
remaining part of the potential (we are ruling out the occurrence of
zero-modes, i.e. non-vanishing eigenvectors $\varepsilon_{\lambda}$
belonging to the zero eigenvalue $\lambda=0$). Thus, the boundary
conditions (1.6) and (1.7) are both gauge invariant under the same
condition on the ghost if Eq. (1.8) is satisfied.
\vskip 0.3cm
\leftline {\bf Acknowledgments}
\vskip 0.3cm
\noindent
Correspondence with Gerd Grubb has been very helpful.
This work has been partially supported by PRIN97 ``Sintesi.'' 
\vskip 1cm
\leftline {\bf References}
\vskip 0.3cm
\item {[1]}
M. Schr\"{o}der, {\it Rep. Math. Phys.} {\bf 27}, 259 (1989).
\item {[2]}
B. S. DeWitt, in {\it Relativity, Groups and Topology II}, eds.
B. S. DeWitt and R. Stora (North Holland, Amsterdam, 1984).
\item {[3]}
G. Esposito, A. Yu. Kamenshchik and G. Pollifrone,
{\it Euclidean Quantum Gravity on Manifolds with Boundary},
Fundamental Theories of Physics, Vol. 85 (Kluwer, Dordrecht,
1997).
\item {[4]}
P. B. Gilkey, {\it Invariance Theory, the Heat Equation and
the Atiyah--Singer Index Theorem} (Chemical Rubber Company,
Boca Raton, 1995).
\item {[5]}
G. Grubb, {\it Functional Calculus of Pseudodifferential
Boundary Problems} (Birkh\"{a}user, Boston, 1996).
\item {[6]}
A. N. Kolmogorov and S. V. Fomin, {\it Elements of the Theory
of Functions and Functional Analysis} (Mir, Moscow, 1980).
\item {[7]}
J. W. Moffat, {\it Phys. Rev.} {\bf D41}, 1177 (1990).
\item {[8]}
J. W. Moffat, {\it Phys. Rev.} {\bf D43}, 499 (1991).
\item {[9]}
G. Esposito, {\it Class. Quantum Grav.} {\bf 16},
1113 (1999).
\item {[10]}
G. Esposito, ``New Kernels in Quantum Gravity'' (HEP-TH 9906169).

\bye